\documentclass[aps,prc,twocolumn,superscriptaddress,preprintnumbers,showpacs,floatfix,nofootinbib]{revtex4-1}
\usepackage{amsmath,graphicx,color,ulem}
\usepackage{hyperref,cleveref}
\newcommand{\trento}{T$\mathrel{\protect\raisebox{-2.1pt}{R}}$ENTo}

\begin{document}

\title{Bayesian reconstruction of anisotropic flow fluctuations at fixed impact parameter}

\author{Enak Roubertie}
\affiliation{Institut de physique th\'eorique, Universit\'e Paris Saclay, CNRS, CEA, F-91191 Gif-sur-Yvette, France}
\affiliation{Mines Paris -- PSL Research University, 60 Boulevard Saint-Michel, 75272 Paris, France}

\author{Mathis Verdan}
\affiliation{Institut de physique th\'eorique, Universit\'e Paris Saclay, CNRS, CEA, F-91191 Gif-sur-Yvette, France}
\affiliation{Mines Paris -- PSL Research University, 60 Boulevard Saint-Michel, 75272 Paris, France}

\author{Andreas Kirchner}
\affiliation{Department of Physics, Duke University, Durham, NC 27708, USA}

\author{Jean-Yves Ollitrault}
\affiliation{Institut de physique th\'eorique, Universit\'e Paris Saclay, CNRS, CEA, F-91191 Gif-sur-Yvette, France}

\date{\today}

\begin{abstract}
  The cumulants of the distribution of anisotropic flow are measured accurately in Pb+Pb collisions at the LHC  as a function of centrality classifiers (charged multiplicity and/or transverse energy).
  Using Bayesian inference, we reconstruct from these measurements the probability distribution of anisotropic flow in the ``theorists' frame'' where the impact parameter has a fixed magnitude and orientation, up to $\sim 70\%$ centrality.
  The variation of flow fluctuations with impact parameter displays direct evidence of viscous damping, which is larger for higher Fourier harmonics, in line with expectations from hydrodynamics. 
  We use intensive measures of non-Gaussian flow fluctuations, which have reduced dependence on centrality. 
  We infer from ATLAS data the magnitude of these intensive non-Gaussianities in each Fourier harmonic. 
  They provide data-driven estimates of response coefficients to initial anisotropies, without resorting to any specific microscopic model of initial conditions.
  These estimates agree with viscous hydrodynamic calculations. 
\end{abstract}
\maketitle

\section{Introduction}
Nucleus-nucleus collisions at ultrarelativistic energies create a droplet of strongly-interacting fluid which expands into the vacuum~\cite{Romatschke:2007mq,Heinz:2013th,Busza:2018rrf}.
The expansion is driven by pressure gradients~\cite{Ollitrault:1992bk}, which are determined by the geometry of the collision volume. 
The impact parameter, defined as the distance between the centers of the two nuclei at the time of impact, plays a crucial role in defining this geometry. 
Now, the experimentally-defined centrality, which is typically based on a classification of events according to multiplicity~\cite{ALICE:2013hur},  is not equivalent to the true centrality, defined according to the impact parameter of the collision. 
For a fixed value of the multiplicity, the true centrality spans a finite range~\cite{Das:2017ned}. 
In recent years, the relevance of these impact parameter fluctuations has been gradually revealed through a number of phenomena occurring in ultra-central Pb+Pb collisions at the Large Hadron Collider (LHC)~\cite{Luzum:2012wu,CMS:2013bza}: 
\begin{itemize}
\item
Rise of mean transverse momentum~\cite{Gardim:2019brr,CMS:2024sgx,Rueda:2024tzc,ATLAS:2024jvf}. 
\item
  Ultracentral flow puzzle, i.e., magnitude of triangular flow, $v_3$, relative 
  to elliptic flow, $v_2$~\cite{Zakharov:2021lux}. 
\item
Decrease of the variance~\cite{ATLAS:2019pvn,Samanta:2023amp,ALICE:2024apz}  of transverse momentum fluctuations.
\item
Increase of the skewness~\cite{Samanta:2023kfk,ALICE:2023tej,ATLAS:2024jvf} of transverse momentum fluctuations.
\item
Positive sign of the 4-particle cumulant of elliptic flow fluctuations, $c_2\{4\}$~\cite{Zhou:2018fxx,ATLAS:2019peb,Alqahtani:2024ejg}.
\end{itemize}
All these phenomena are quantitatively explained, provided that one properly takes into account the fluctuations of impact parameter relative to the centrality classifier, i.e., the observable used by experiments to define centrality classes. 
State-of-the-art models~\cite{Bernhard:2016tnd,Nijs:2020roc,JETSCAPE:2020mzn} do include such impact parameter fluctuations. 
However, we will argue that their magnitude is unlikely to be right, because these models are calibrated using a set of experimental data which does not contain the relevant information. 

Dedicated studies of impact parameter fluctuations have been limited to central collisions (typically $0-10\%$)  so far.  
We extend this study all the way to 70\% centrality, focusing on anisotropic flow data in Pb+Pb collisions at $\sqrt{s_{\rm NN}}=5.02$~TeV published by the ATLAS Collaboration~\cite{ATLAS:2019peb}.
The centrality resolution becomes significantly worse as the centrality percentile increases. 
Therefore, it is important to assess quantitatively the effects of impact parameter fluctuations, in particular for elliptic flow, which is mostly driven by the almond shape of the overlap area between the colliding nuclei, and depends strongly on impact parameter~\cite{Ollitrault:1992bk}.

In Sec.~\ref{s:2intro}, we briefly recall how final-state observables are related to the initial stages of the collision~\cite{Luzum:2013yya} through the hydrodynamic evolution.
We explain why it is important to distinguish impact parameter fluctuations, which are classical fluctuations, from the remaining fluctuations, which are quantum. 

Our study focuses on collisions between spherical nuclei. 
We show that the probability distribution of anisotropic flow at fixed impact parameter can be parametrized in a robust way, without resorting to specific microscopic models.
It is to a good approximation Gaussian~\cite{Voloshin:2007pc}, and corrections to the Gaussian can be classified systematically using a cumulant expansion~\cite{Giacalone:2016eyu,Abbasi:2017ajp}, as recalled in Sec.~\ref{s:cumulants}.
We identify three leading corrections to the naive Gaussian model of elliptic flow fluctuations:
Fluctuation asymmetry, skewness and kurtosis, which we loosely refer to as non-Gaussianities. 
We introduce {\it intensive\/} measures of non-Gaussianities~\cite{Alqahtani:2024ejg}, in the sense that their system-size dependence is expected to be mild. 
They generalize the intensive skewness $\Gamma_{p_t}$ of transverse momentum fluctuations~\cite{Giacalone:2020lbm,ALICE:2023tej,ATLAS:2024jvf}. 
We  show that intensive non-Gaussianities are significantly increased by the hydrodynamic response to initial anisotropies~\cite{Gardim:2011xv,Teaney:2012ke}. 

In Sec.~\ref{s:centrality}, we reconstruct the probability distribution of the true centrality at fixed ``experimental'' centrality.  
This reconstruction is robust in central collisions~\cite{Das:2017ned}, where the resolution is inferred from the tail of the multiplicity distribution. 
How the resolution varies with impact parameter cannot be inferred from data alone.
We argue that it can be determined using standard models of initial conditions, and that this determination is robust with respect to model details. 
We show in particular that fluctuations are suppressed in central collisions, relative to the naive expectation based on independent sources.
This phenomenon can be traced back to the number of participant nucleons, whose fluctuations undergo strong binomial suppression in central collisions. 

In Sec.~\ref{s:fits}, we fit anisotropic flow data on the cumulants of elliptic ($v_2$), triangular ($v_3$) and quadrangular ($v_4$) flow  from the ATLAS Collaboration ~\cite{ATLAS:2019peb}, which are measured as a function of two different centrality classifiers, the charged multiplicity around mid-rapidity $N_{ch}$ and the transverse energy at large rapidity $E_T$. 
We extend the work initiated in Ref.~\cite{Alqahtani:2024ejg}, which was limited to the 5\% most central events, up to 70\% centrality.

Results are discussed in Sec.~\ref{s:interpretation}. 
We unfold the effects of impact parameter fluctuations and achieve a precise reconstruction of the centrality dependence of the mean elliptic flow in the reaction plane, and of $v_2$ and $v_3$ fluctuations.
We isolate the nonlinear contribution to $v_4$, induced by $v_2$, from the residual contribution due to fluctuations.
We obtain in particular robust estimates of  intensive non-Gaussianities in all harmonics, which yield quantitative information about the hydrodynamic response. 

\section{Sources of flow fluctuations}
\label{s:2intro}

\subsection{Hydrodynamic response to initial anisotropies}
\label{s:hydroresponse}

In the hydrodynamic picture of heavy-ion collisions~\cite{Romatschke:2017ejr}, final-state observables are determined by the energy density profile right after the collision.
Detailed simulations have shown that this relation between initial and final state can be encapsulated, to a good approximation, into simple response relations:
Our study focuses on anisotropic flow, which we write in complex notation as $v_n\equiv\langle e^{in\varphi}\rangle$, where the average is over outgoing particles with azimuthal angles $\varphi$.
For elliptic ($n=2$) and triangular ($n=3$) flows, $v_n$ is determined to a good approximation by linear response to the complex anisotropy $\varepsilon_n$, defined as a Fourier coefficient of the initial density profile~\cite{PHOBOS:2006dbo,Qiu:2011iv,Teaney:2010vd}:
\begin{equation}
\label{linear}
v_n=\kappa_n\varepsilon_n,
\end{equation}
where $\kappa_n$ is a real response coefficient~\cite{Gardim:2011xv,Niemi:2012aj},\footnote{For elliptic flow at large impact parameters, $v_2$ increases slightly faster than $\varepsilon_2$~\cite{Niemi:2015qia}, which has been attributed to a small cubic response~\cite{Noronha-Hostler:2015dbi}. We neglect this nonlinearity throughout this paper.}  and $\varepsilon_n$ lies within the unit disk, $|\varepsilon_n|<1$, a property that will prove crucial below.
The response coefficient only depends on kinematic cuts of the experiment (in particular the cuts in transverse momentum $p_t$) and of the properties of the fluid (equation of state and transport coefficients~\cite{Gardim:2020mmy}). 
The event-by-event fluctuations of $v_n$ solely originate from those of $\varepsilon_n$.

Higher harmonics ($n\ge 4$) involve elliptic flow through nonlinear response terms~\cite{Borghini:2005kd,Gardim:2011xv} and are not as well understood.
One usually assumes that $v_4$ can be decomposed as the sum of two independent terms
\begin{equation}
\label{nonlinear}
v_4=\chi_{4,22} (v_2)^2+\kappa_4 {\cal C}_4. 
\end{equation}
The first term is the nonlinear response, defined by a dimensionless coefficient $\chi_{4,22}$~\cite{Qian:2016fpi}.
The second term is determined by linear response to a dimensionless complex coefficient ${\cal C}_4$~\cite{Teaney:2012ke}, which is a cumulant involving the fourth Fourier harmonic of the initial density profile $\varepsilon_4$, corrected to subtract the contribution of the second harmonic. 
We further simplify this picture by assuming that the two terms are independent~\cite{Yan:2015jma,ALICE:2017fcd,ALICE:2019xkq,CMS:2019nct,ALICE:2020sup,STAR:2020gcl,STAR:2022vkx}.

A long-standing problem in heavy-ion phenomenology is that from data alone, it is very hard to constrain separately the hydrodynamic response $\kappa_n$ and the initial anisotropy  $\varepsilon_n$ (or $ {\cal C}_4$ for $n=4$). 
A larger anisotropy can typically be compensated by a larger viscosity, which reduces the hydrodynamic response~\cite{Luzum:2008cw}. 
We will suggest a way to circumvent this problem by scrutinizing the probability distribution of flow fluctuations. 

\subsection{Classical versus quantum fluctuations}

Since the event-by-event fluctuations of $v_n$ result from those of $\varepsilon_n$, they can be traced back to fluctuations of the initial density profile, which have two distinct sources. 

First, the impact parameter is not identical for all events. 
It is a classical quantity~\cite{Samanta:2023amp}, in the sense that the quantum uncertainties on its magnitude and direction are negligible at LHC energies. 
The size and shape of the collision volume, which determines the properties of outgoing particles, depend strongly on the impact parameter, so that this classical fluctuation has a non-trivial effect. 

Our main point is that once the impact parameter is fixed, the remaining event-by-event fluctuations are simpler.
They result from quantum fluctuations in the wavefunctions of incoming nuclei or in the collision process. 
Due to these quantum fluctuations, the energy density profile formed just after the collision is not smooth but presents hot spots, which are local in the transverse plane and elongated along the longitudinal direction~\cite{Takahashi:2009na,Schenke:2010rr}. 
The crucial point is that in a collision between two large nuclei, fluctuations occurring in different points of the transverse plane are to a large extent independent~\cite{Blaizot:2014nia} as they are causally disconnected.\footnote{This only holds for spherical nuclei~\cite{Giacalone:2023hwk}. For deformed nuclei, event-by-event fluctuations in orientation induce a global modification of the density profile~\cite{Heinz:2004ir}, i.e., they induce long-range correlations. 
They are quasi-classical fluctuations, much in the same way as impact parameter fluctuations. 
Note also that even for spherical nuclei, there is a long-range correlation due to the fact that the number of nucleons is fixed, so that more participant nucleons at one point implies fewer elsewhere. We come back to this effect in Sec.~\ref{s:binomial}.}
Therefore, the fluctuations of quantities involving integrals over the transverse plane, such as $\varepsilon_n$ or ${\cal C}_4$~\cite{PHOBOS:2006dbo,Teaney:2010vd}, are constrained by the central limit theorem, and are approximately Gaussian. 
We will use this property in Sec.~\ref{s:cumulants}. 

\section{Anisotropic flow fluctuations in the intrinsic frame}
\label{s:cumulants}

We carry out a thought experiment where the magnitude and direction of impact parameter are identical for all collisions.
We refer to this theorist's view as the intrinsic frame. 
As explained above, fluctuations in the intrinsic frame are independent quantum fluctuations which are constrained by the central limit theorem. 
The Gaussian model of eccentricity fluctuations~\cite{Voloshin:2007pc} is recalled in Sec.~\ref{s:ecccumulants}, and the generic next-to-leading corrections to this Gaussian model are listed: 
Asymmetry, skewness, kurtosis. 
In Sec.~\ref{s:intensive}, we introduce intensive measures of these non-Gaussianities, which have reduced dependence on the system size. 
They are evaluated explicitly in Sec.~\ref{s:ellipticpower} for the elliptic-power distribution, which is an analytic toy model of eccentricity fluctuations~\cite{Yan:2014afa}. 
In Sec.~\ref{s:response}, we show that intensive non-Gaussianities are enhanced by the hydrodynamic response. 

\subsection{Standard model of Gaussian eccentricity fluctuations, and first corrections to this model} 
\label{s:ecccumulants}

Monte Carlo simulations implementing standard models of initial conditions show that event-by-event fluctuations of 
$\varepsilon_2$ in the intrinsic frame are approximately Gaussian~\cite{Voloshin:2007pc} for spherical nuclei. 
The probability density of $\varepsilon_2$  is of the form
\begin{equation}
\label{gaussian}
P(\varepsilon_2)=\frac{1}{\pi\sigma_{\varepsilon_2}^2}\exp\left(-\frac{\left|\varepsilon_2-\bar\varepsilon_2\right|^2}{\sigma_{\varepsilon_2}^2}\right).
\end{equation}
It is specified by its mean $\bar\varepsilon_2$ and standard deviation $\sigma_{\varepsilon_2}$:
\begin{align}
\langle\varepsilon_2\rangle &=\bar\varepsilon_2\nonumber\\
\langle\varepsilon_2\varepsilon_2^*\rangle- \langle\varepsilon_2\rangle\langle\varepsilon_2^*\rangle&= \sigma_{\varepsilon_2}^2,
\label{firstcumulants}
\end{align}
where $\varepsilon_2^*$ is the complex conjugate of $\varepsilon_2$, and angular brackets denote the average with the probability distribution $P(\varepsilon_2)$. 
We follow the standard convention where the $x$ axis is the direction of the impact parameter, also referred to as ``reaction plane'' (which is, more precisely, the $(x,z)$ plane). 
In this coordinate system, the symmetry of the probability distribution with respect to the reaction plane is $y\to -y$ symmetry. 
It implies that $\bar\varepsilon_2$ is real. 

The mean and the standard deviation are the lowest-order cumulants. 
We define the cumulant $c_{mp}$ by~\cite{Mehrabpour:2020wlu,Alqahtani:2024ejg}
\begin{equation}
  \label{defcmp}
c_{mp}\equiv\langle (\varepsilon_2^*)^m(\varepsilon_2)^p\rangle_c, 
\end{equation}
where the subscript $c$ indicates that one isolates the connected part by subtracting out disconnected components order by order, as in the second line of Eq.~(\ref{firstcumulants}). 
The cumulants $c_{mp}$ are the elements of an infinite matrix, whose first element $c_{00}$ vanishes by convention. The $y\to -y$  symmetry implies a $\varepsilon_2\to \varepsilon_2^*$ symmetry, so that the matrix is real and symmetric, $c_{mp}=c_{mp}^*=c_{pm}$. 

With this notation, Eq.~(\ref{firstcumulants}) can be rewritten as: 
\begin{align}
c_{01} &=\bar\varepsilon_2\nonumber\\
c_{11}&=\sigma_{\varepsilon_2}^2. \label{firstcumulantsbis}
\end{align}
For the symmetric Gaussian distribution (\ref{gaussian}), all other cumulants vanish, and $c_{mp}$ reduces to a $2\times 2$ matrix. 
This distribution only represents a leading approximation, which can be systematically improved by including higher-order cumulants. 
The next-to-leading corrections are obtained by including cumulants with $p=2$, so that $c_{mp}$ is promoted to a $3\times 3$ matrix. 
The three new cumulants are: 
\begin{align}
c_{02} &=\langle (\varepsilon_2- \bar\varepsilon_2)^2\rangle=\langle (\varepsilon_{2x}- \bar\varepsilon_2)^2\rangle-\langle\varepsilon_{2y}^2\rangle\nonumber\\
c_{12}&= \langle (\varepsilon_2^*-\bar\varepsilon_2)(\varepsilon_2-\bar\varepsilon_2)^2\rangle\nonumber\\
c_{22}&= \langle (\varepsilon_2^*-\bar\varepsilon_2)^2(\varepsilon_2-\bar\varepsilon_2)^2\rangle-2\sigma_{\varepsilon_2}^4-c_{02}^2.
\label{nextcumulants}
\end{align}
Their physical interpretation is the following. 
$c_{02}$ represents the difference between the magnitudes of eccentricity fluctuations parallel ($x$) and perpendicular ($y$) to the reaction plane, that is, the asymmetry of eccentricity fluctuations. 
$c_{12}$ is a centered moment or order 3, which quantifies the skewness of the eccentricity distribution~\cite{Giacalone:2016eyu,CMS:2017glf,ALICE:2018rtz}, while $c_{22}$ is its excess kurtosis~\cite{Bhalerao:2018anl,CMS:2023bvg}, which we refer to simply as kurtosis.  

Our study is limited to these next-to-leading corrections to the Gaussian model. 
It could be improved by adding next-to-next-to-leading corrections, corresponding to cumulants with $p=3$. 
The superskewness of $v_2$ fluctuations (introduced by CMS in Ref.~\cite{CMS:2023bvg}, where it is denoted by $p_{50}$), which will not be studied here, is one such correction, which corresponds to the cumulant $c_{23}$ in our notation (see Appendix \ref{s:cversusk}).  

The same discussion applies to the fluctuations of $\varepsilon_3$, with one important simplification: 
For symmetric collisions near mid-rapidity, the probability distribution of energy density is symmetric under the exchange of target and projectile, which amounts to $\phi\to\phi+\pi$ symmetry. 
This symmetry implies that the average asymmetry $\bar\varepsilon_3$ vanishes:
Triangular flow is solely due to fluctuations~\cite{Alver:2010gr}, whose width $\sigma_{\varepsilon_3}$ may differ from $\sigma_{\varepsilon_2}$. 
The same symmetry argument implies that the skewness vanishes, $c_{12}=0$. 
A non-zero fluctuation asymmetry $c_{02}$ is allowed, but it involves a sixth Fourier harmonic of the energy profile, and we shall neglect it. 
The only correction to the Gaussian model which we retain is the kurtosis, $c_{22}$. 
Finally, we will assume for simplicity that the distribution of ${\cal C}_4$ in Eq.~(\ref{nonlinear}) can be modeled in the same way as the distribution of $v_3$: 
A symmetric Gaussian, corrected by a small kurtosis. 

\subsection{Intensive cumulants of eccentricity fluctuations}
\label{s:intensive}

We now derive the order of magnitude of $c_{mp}$~\cite{Alqahtani:2024ejg}. 
For $N$ independent sources, with $N\gg 1$, $c_{mp}$ is proportional to $N^{1-m-p}$: Adding one power of $\varepsilon$ or $\varepsilon^*$ adds a factor $1/N$~\cite{Bhatta:2021qfk}. 
Additional information is provided by the azimuthal structure. 
Writing $\varepsilon_2=|\varepsilon_2| e^{2i\phi}$, one sees that the quantity inside brackets in Eq.~(\ref{defcmp}) involves a factor $e^{2i(p-m)\phi}$. 
For central collisions, azimuthal symmetry then implies that the only non-vanishing cumulants are the diagonal ones, $c_{mm}$. 
For off-central collisions, $c_{mp}$ is a Fourier coefficient of the eccentricity distribution  of order $2|p-m|$, which typically scales like $\bar\varepsilon_2^{|p-m|}$. 
In order to eliminate the trivial dependence of cumulants on $N$ and $\bar\varepsilon_2$, we define the {\it intensive\/}~\cite{Giacalone:2020lbm,ALICE:2023tej,ATLAS:2024jvf} cumulants $\Gamma_{mp}$ by 
\begin{equation}
\label{defintensive}
\Gamma_{mp}\equiv \frac{c_{mp}}{\sigma_{\varepsilon_2}^{2(m+p-1)}\bar\varepsilon_2^{|p-m|}}=\frac{c_{mp}}{c_{11}^{m+p-1}c_{01}^{|p-m|}}.
\end{equation}
Note that $\Gamma_{01}=\Gamma_{11}=1$ by construction, so that the only non-trivial intensive cumulants with $m\le p$ are those with $p\ge 2$.  
One expects that $\Gamma_{mp}$ depends much less on the anisotropy and size of the source than $c_{mp}$, as will be shown in Sec.~\ref{s:ellipticpower}. 

As explained in Sec.~\ref{s:ecccumulants}, we only retain cumulants with $p=2$.
The associated intensive cumulants are the intensive asymmetry $\Gamma_{02}$, the intensive skewness $\Gamma_{12}$ and the intensive kurtosis $\Gamma_{22}$.

\subsection{An explicit example: The elliptic-power distribution} 
\label{s:ellipticpower}

We now illustrate the general properties of cumulants on the example of the elliptic-power distribution. 
We argue that the intensive cumulants of this distribution, which can be calculated analytically, are to a large extent universal, and that they should be approximately the same for any realistic model of initial conditions. 
\begin{figure}[ht]
\begin{center}
\includegraphics[width=.7\linewidth]{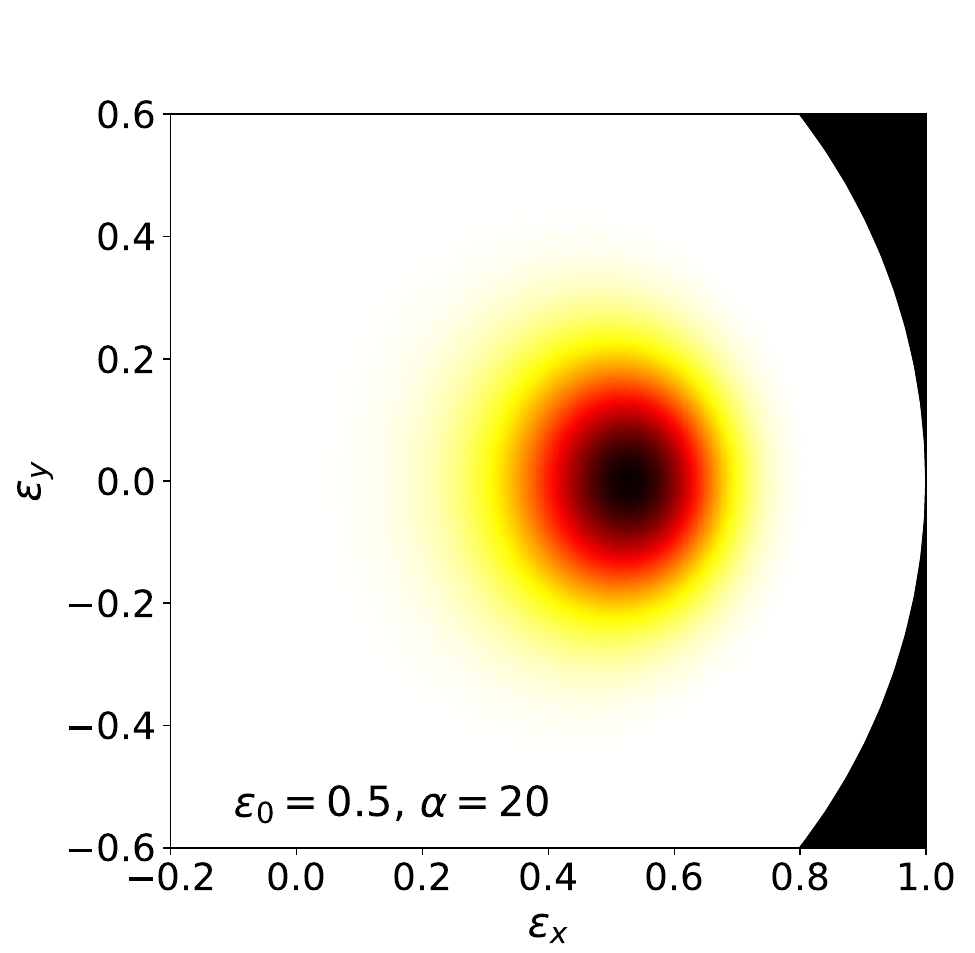} 
\includegraphics[width=.7\linewidth]{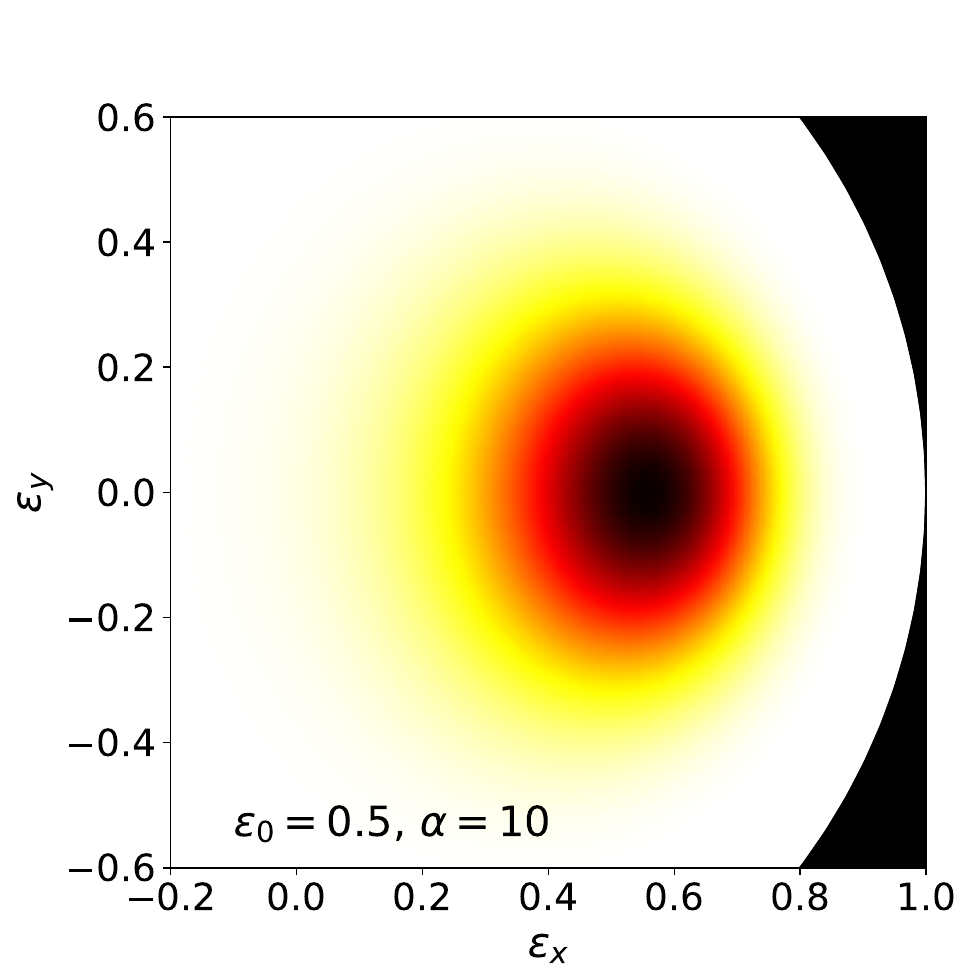}
\end{center}
\caption{ 
  \label{fig:ellipticpower}
Representation of the elliptic-power distribution (\ref{ellipticpower2d}). 
The forbidden region $\varepsilon_x^2+\varepsilon_y^2>1$ is represented in black color. 
 }
\end{figure}    

The elliptic-power distribution is the {\it exact\/} distribution of $\varepsilon_2$ for a density profile made of $N$ independent and identical point-like sources with a Gaussian distribution in the transverse plane~\cite{Ollitrault:1992bk}. 
Its explicit expression is~\cite{Yan:2014afa}
\begin{equation}
P(\varepsilon_x,\varepsilon_y)=
\frac{\alpha}{\pi}
  (1-\varepsilon_0^2)^{\alpha+\frac{1}{2}}\frac{(1-\varepsilon_x^2-\varepsilon_y^2)^{\alpha-1}}
{(1-\varepsilon_0\varepsilon_x)^{2\alpha+1}},
\label{ellipticpower2d}
\end{equation}
where $\varepsilon_x$ and $\varepsilon_y$ denote the real and imaginary parts of $\varepsilon_2$,   $\alpha=(N-2)/2$\footnote{One must evaluate the center of the energy density profile on an event-by-event basis before evaluating the eccentricity with respect to this center~\cite{PHOBOS:2006dbo}. This recentering correction amounts to replacing $N$ with $N-1$ in the elliptic-power distribution~\cite{Gronqvist:2016hym}.} and $\varepsilon_0$ parametrizes the average value of $\varepsilon_x$. 
The support of $P(\varepsilon_x,\varepsilon_y)$ is the unit disk, $\varepsilon_x^2+\varepsilon_y^2\le 1$. 
This is implied by the definition of $\varepsilon_2$, which is a scaled Fourier coefficient so that $|\varepsilon_2|\le 1$.

Fig.~\ref{fig:ellipticpower} displays the elliptic-power distribution for $\varepsilon_0=0.5$ and two different values of $\alpha$. 
Its center lies approximately at $(0.5,0)$, which illustrates that the average eccentricity in the reaction plane, $\bar\varepsilon_2$, is close to $\varepsilon_0$. 
The distribution is more elongated along the $y$-axis, which means that the asymmetry $c_{02}$ defined by Eq.~(\ref{nextcumulants})  is negative. 
In addition, it has a longer tail to the left than to the right, which means that it has negative skew, i.e., $c_{12}$ is also negative. 
These non-Gaussian features are less prominent for the larger values of $\alpha$, as expected from the central limit theorem. 
It is intuitively obvious that both features result from the condition $|\varepsilon_2|\le 1$. 
This boundary drives both the asymmetry and the skewness of the distribution.

\begin{figure}[ht]
\begin{center}
\includegraphics[width=.8\linewidth]{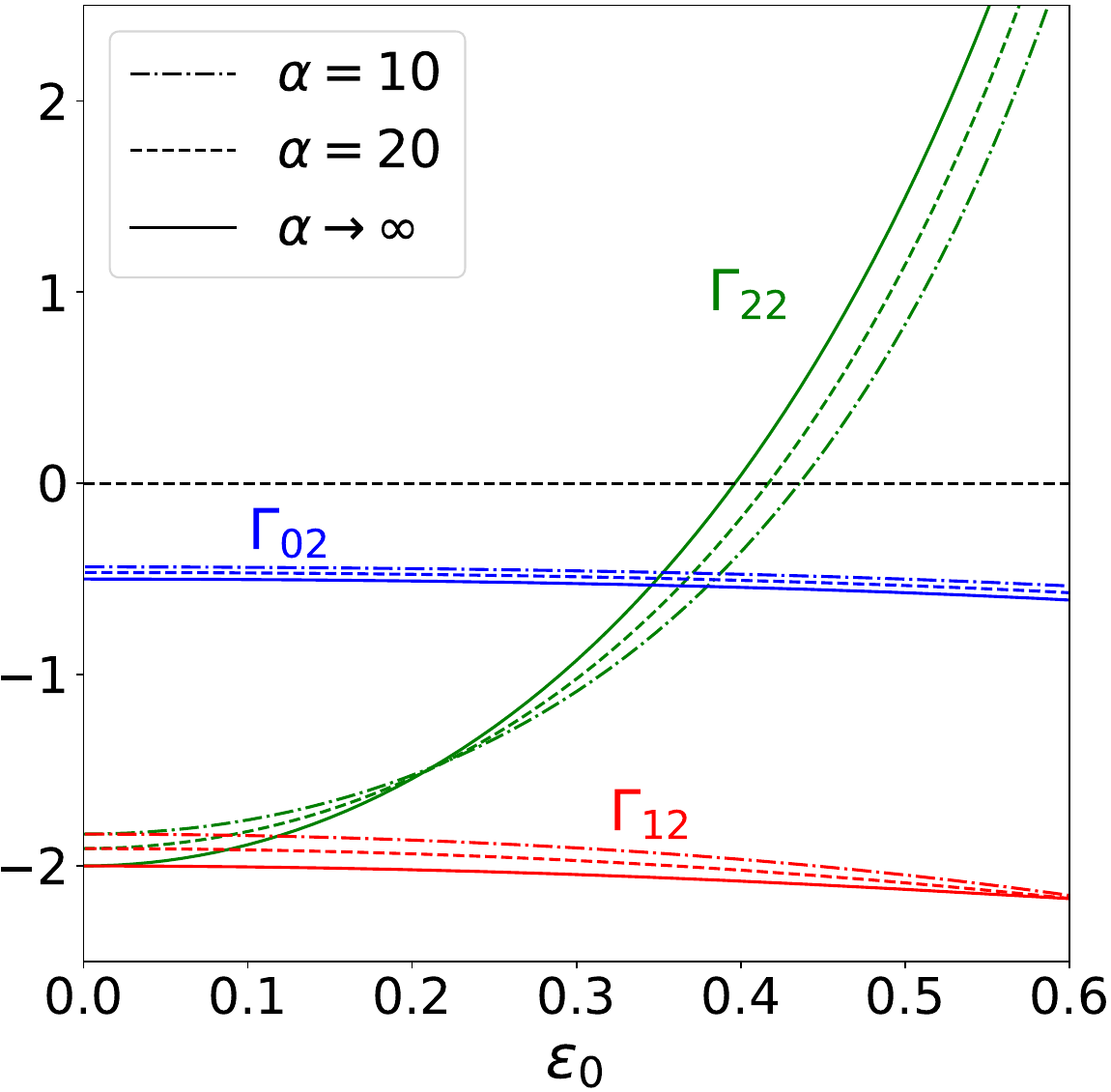} 
\end{center}
\caption{ 
  \label{fig:epintensive}
Intensive asymmetry $\Gamma_{02}$, skewness $\Gamma_{12}$ and kurtosis $\Gamma_{22}$ of the elliptic-power distribution as a function of the average anisotropy $\varepsilon_0$. 
 }
\end{figure}    

The cumulants of the elliptic-power distribution can be calculated analytically~\cite{Abbasi:2017ajp}. 
Fig.~\ref{fig:epintensive} displays the variation of the intensive cumulants $\Gamma_{02}$, $\Gamma_{12}$ and $\Gamma_{22}$ as a function of $\varepsilon_0$ for fixed $\alpha$. 
One sees that they depend little on $\alpha$ in the limit $\alpha\gg 1$. 
Their asymptotic expressions are:
\begin{align}
\Gamma_{02}&= -\frac{1}{2-\epsilon_0^2}\nonumber\\
\Gamma_{12}&= -2\frac{4-3\epsilon_0^2}{(2-\epsilon_0^2)^2}\nonumber\\
\Gamma_{22}&= -2\frac{8-64\epsilon_0^2+89\epsilon_0^4-36\epsilon_0^6}{(1-\epsilon_0^2)(2-\epsilon_0^2)^3}. 
\label{epintensive}
\end{align}
For realistic values of $\varepsilon_0$, which is not expected to exceed $0.6$, the intensive asymmetry and intensive skewness vary little, and are approximately equal to $-\frac{1}{2}$ and $-2$, respectively. 
The intensive kurtosis, on the other hand, varies strongly with $\varepsilon_0$, and changes sign from negative to positive as $\varepsilon_0$ increases. 
The negative sign for small $\varepsilon_0$ is again a natural consequence of the boundary condition $|\varepsilon_2|\le 1$, which forces the distribution to be narrower than a Gaussian~\cite{Yan:2013laa}.

Even though the elliptic-power distribution is a toy model, its qualitative features are generic. 
Realistic models of initial conditions also predict a negative skewness of $v_2$ fluctuations (stemming from that of $\varepsilon_2$ fluctuations)~\cite{Giacalone:2016eyu,Abbasi:2017ajp}, which has been confirmed experimentally~\cite{CMS:2017glf,ALICE:2018rtz}.
The change of sign of the kurtosis from negative to positive has also been predicted~\cite{Bhalerao:2018anl} and observed~\cite{CMS:2023bvg}. 
As for triangular flow, the distribution of $\varepsilon_3$ has been shown to be well approximated by the elliptic-power distribution with $\varepsilon_0=0$, also referred to as the power distribution~\cite{Yan:2013laa}. 
The negative kurtosis is the origin of the non-zero $v_3\{4\}$ which has long been observed~\cite{ALICE:2011ab}, and then found to be in quantitative agreement with hydrodynamic simulations~\cite{Giacalone:2017uqx}. 

We will postulate in our analysis that the intensive asymmetry and skewness of the distribution of $\varepsilon_2$ are approximately equal to $-\frac{1}{2}$ and $-2$ respectively, corresponding to the value for the elliptic-power distribution for moderate $\varepsilon_0$,  and that the intensive kurtosis of $\varepsilon_3$ and ${\cal C}_4$ is approximately equal to $-2$, corresponding to the value for the elliptic-power distribution for $\varepsilon_0=0$. 
These assumptions should eventually be confirmed by state-of-the-art simulations of the initial density profile. 
Note that detailed comparisons between the elliptic-power distributions and realistic initial condition models have already been carried  out~\cite{Abbasi:2017ajp}, but with a centrality selection done as in experiment. 
We expect that agreement would be improved by running the models at fixed impact parameter. 
This is beyond the scope of the present article and will be carried out in a future publication. 

\subsection{Intensive cumulants of flow fluctuations}
\label{s:response}

The cumulants of the distribution of $v_n$ in the intrinsic frame can be defined in the same way as those of $\varepsilon_n$, by replacing  $\varepsilon_2$ with the complex $v_n$ in Eq.~(\ref{defcmp}). 
Under the assumption of linear hydrodynamic response, Eq.~(\ref{linear}) gives immediately
\begin{equation}
\label{cmpv}
c_{mp}^{(v_n)}=\kappa_n^{m+p} c_{mp}^{(\varepsilon_n)}. 
\end{equation}
Injecting in Eq.~(\ref{defintensive}), one obtains that the intensive cumulants of the $v_n$ distribution are related to those of the $\varepsilon_n$ distribution through: 
\begin{equation}
\label{intensivev}
\Gamma_{mp}^{(v_n)}=\frac{1}{\kappa_n^{2(p-1)}} \Gamma_{mp}^{(\varepsilon_n)}
\end{equation}
for $m\le p$. 

Thus the intensive cumulants of $v_2$ associated with the next-to-leading corrections to the Gaussian model, Eq.~(\ref{nextcumulants}), which correspond to $p=2$, are enhanced by a factor $1/\kappa_2^2$ relative to those of $\varepsilon_2$ fluctuations, since the response coefficient is smaller than unity~\cite{Kolb:2000sd}. 
Similarly, the kurtosis of $v_3$ and $v_4$ fluctuations are enhanced by a factor $1/\kappa_3^2$ and $1/\kappa_4^2$~\cite{Alqahtani:2024ejg}. 

As will be illustrated in Sec.~\ref{s:fitvsdata}, intensive cumulants are more convenient measures of non-Gaussianities than the standardized skewness~\cite{Giacalone:2016eyu,Abbasi:2017ajp,CMS:2017glf,ALICE:2018rtz} and standardized kurtosis~\cite{Bhalerao:2018anl,CMS:2023bvg} which have been used so far.
Standardized cumulants are devised in such a way that they are smaller for large systems, so that they have a strong centrality dependence. 
This dependence is reduced for intensive cumulants.

\section{Centrality fluctuations}
\label{s:centrality}

The experimental analysis is done for a sample of events which spans a range of impact parameters.
We now explain how this range can be evaluated quantitatively. 

In order to relate the experimental centrality $c_{\rm exp}$ to the true centrality $c$, we use the Bayesian method introduced in Ref.~\cite{Das:2017ned}.
It relies on the assumption that fluctuations of the centrality classifier (charged multiplicity $N_{ch}$ or transverse energy $E_T$ in the case of ATLAS~\cite{ATLAS:2019peb}) are Gaussian at fixed $c$.
We denote generically by $N$ this centrality classifier.
The reconstruction boils down to determining its mean $\overline{N}$ and variance $\sigma_N^2$ as a function of $c$.
By fitting the distribution of $N$, one can determine $\overline{N}(c)$ and $\sigma_N^2(0)$.
There is at present no way of determining the centrality dependence of $\sigma_N^2$ from experimental data.

This implies that the width of centrality fluctuations, which depend on the relative multiplicity fluctuation $\sigma_N(c)/\overline{N}(c)$, is known only for central collisions.
This is the reason why previous studies of centrality fluctuations~\cite{Samanta:2023amp,Alqahtani:2024ejg} have only addressed a limited range of centrality. 

In order to extend this range, we resort to microscopic models of the collision in order to gather some information about the centrality dependence of fluctuations. 
More specifically, we assume that $N$ is proportional to the entropy produced in the early stages of the collision on an event-by-event-basis, and we evaluate this entropy using the \trento{} model of initial conditions~\cite{Moreland:2014oya}

\subsection{Binomial suppression of fluctuations} 
\label{s:binomial}

\trento{} is a versatile model of initial conditions, which is widely used in hydrodynamic calculations~\cite{Bernhard:2016tnd,Nijs:2020roc,JETSCAPE:2020mzn}.
It returns for each event an entropy density profile in the transverse plane $s(x,y)$. 
The simulation starts by sampling the positions of nucleons within incoming nuclei, and determining which of them participate in the Pb+Pb collision, on the basis of the Glauber model~\cite{Miller:2007ri}. 
Then, there are free parameters which determine the proton width $w$, the number of constituents $n_c$ in each participant nucleon~\cite{Moreland:2018gsh}, their width $v$, and how much the thickness of a given nucleon is allowed to fluctuate (more precisely, the thickness is sampled from a gamma distribution, and the free parameter is the shape parameter of the gamma distribution). 
Thickness functions $T_A(x,y)$ and $T_B(x,y)$ are obtained by summing the thicknesses of participants in each nucleus.
Finally, there is a parameter $p$ which determines how $s$ depends on $T_A$ and $T_B$.
We choose the value $p=0$, corresponding to $s\propto\sqrt{T_AT_B}$, which is favored all by theory-to-data comparisons~\cite{Moreland:2018gsh,JETSCAPE:2020mzn,Nijs:2020roc}. 

The values of other parameters are less constrained by data, and we vary them to assess the robustness of our results. 
The preferred value of the nucleon width has changed~\cite{Giacalone:2022hnz} from $w\simeq 0.5$~fm~\cite{Bernhard:2016tnd}
to $w\simeq 1$~fm~\cite{Moreland:2018gsh,JETSCAPE:2020mzn} and then back to the original value~\cite{Nijs:2022rme}. 
We test both values. 
We also run the model with $n_c=1$ (no nucleonic substructure), and with $n_c=3$ constituents of width $v=0.3$~fm. 

For each set of parameters, we first generate $5\times 10^4$ events with $b=0$. 
We tune the shape parameter of the gamma distribution in such a way that the relative entropy fluctuation in central collisions ($b=0$) matches the relative multiplicity fluctuation inferred from ATLAS data~\cite{Yousefnia:2021cup}, $\sigma_{N_{ch}}(0)/\overline{N_{ch}}(0)=0.045$.
It is interesting to note that state-of-the-art theory-to-data comparisons~\cite{Bernhard:2016tnd,Nijs:2020roc,JETSCAPE:2020mzn} do not include this information in the calibration.
Therefore, there is no guarantee that the magnitude of centrality fluctuations is right.
We then rescale the entropy by a constant factor so that it matches the average multiplicity at $b=0$ in the ATLAS experiment, $\overline{N_{ch}}(0)=3104$~\cite{Yousefnia:2021cup}. 

\begin{figure}[ht]
\begin{center}
\includegraphics[width=\linewidth]{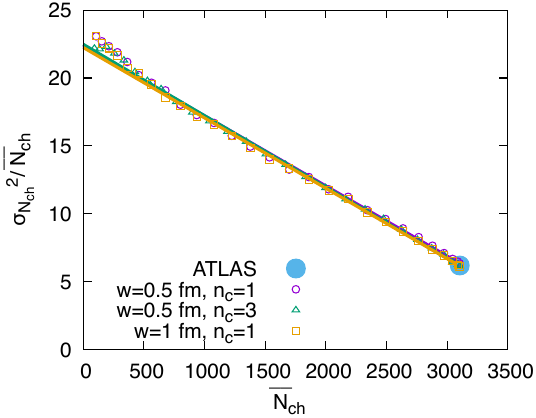}
\end{center}
\caption{ 
  \label{fig:binomial}
  Variation of $\sigma_{N_{ch}}^2/\overline{N_{ch}}$ with $\overline{N_{ch}}$ in \trento{} simulations, for two values of the nucleon width $w$ and the number of constituents $n_c$. 
  Each symbol represents a value of impact parameter (from left to right, $b=13$~fm to $b=0$, by steps of $0.5$~fm). 
  The only data point is at $b=0$ and results from a Bayesian reconstruction~\cite{Yousefnia:2021cup}. 
  Lines are linear fits to each set of symbols, excluding the most peripheral collisions with $N_{ch}<300$.  
 }
\end{figure}    

Then, we run the model for other fixed values of $b$, and generate $5\times 10^4$ events for each $b$. 
We evaluate the mean $\overline{N_{ch}}$ and variance $\sigma_{N_{ch}}^2$ of this sample of events. 
As explained above, they are tuned to ATLAS data for $b=0$.
But they depend somewhat on model parameters for other values of $b$. 
However, when one plots the ratio  $\sigma_{N_{ch}}^2/\overline{N_{ch}}$ as a function of $\overline{N_{ch}}$, all results fall on the same curve, displayed in Fig.~\ref{fig:binomial}. 

One observes that $\sigma_{N_{ch}}^2/\overline{N_{ch}}$ decreases approximately linearly with $\overline{N_{ch}}$, which has already been pointed out~\cite{Jia:2020tvb}, and we now interpret this result. 
If the multiplicity was the superposition of a large number of independent contributions, both the variance and the mean would be proportional to that number, and $\sigma_{N_{ch}}^2/\overline{N_{ch}}$ would be independent of $\overline{N_{ch}}$. 
Our interpretation is that the observed decrease is driven by the fluctuations in the number of participant nucleons, which is determined by the Glauber model underlying \trento{}. 
In collisions at $b=0$, the average number of participants is $N_{\rm part}=410$, very close to the total number of nucleons $2A=416$, which leaves little room for fluctuations. 
The condition $N_{\rm part}<2A$ suggests that the fluctuations of $N_{\rm part}$ at fixed centrality follow a binomial distribution, which implies
\begin{equation}
  \label{binomial}
  \frac{\sigma_{N_{\rm part}}^2}{\overline{N_{\rm part}}}=1-\frac{\overline{N_{\rm part}}}{2A}. 
\end{equation}
Since $N_{ch}$ is in general proportional to $N_{\rm part}$~\cite{Eremin:2003qn,Bialas:2006kw,Bozek:2016kpf}, this linear decrease drives that observed in Fig.~\ref{fig:binomial}, and we therefore refer to this phenomenon as binomial suppression of fluctuations. 
A similar suppression is observed in the context of net-baryon fluctuations~\cite{Bzdak:2012ab,STAR:2013gus,Asakawa:2015ybt,Rustamov:2017lio,Rogly:2018kus}. 

Our interpretation of the observation that models with different parameters fall on the same curve is that there is a common underlying Glauber model. 
Thus we obtain a robust estimate of $\sigma_{N_{ch}}$ for non-central collisions, which is independent of model details. 

We assume for simplicity that the decrease is {\it exactly\/} linear. 
Then, the centrality dependence of $\sigma_N^2$ is encapsulated into a single parameter $\gamma$, defined as the factor by which $\sigma_N^2/\overline{N}$ decreases between peripheral and central collisions~\cite{Samanta:2023kfk}:
\begin{equation}
  \label{defgamma}
  \gamma\equiv\frac{(\sigma^2_N/\overline{N})_{\overline{N}\to 0}}{(\sigma^2_N/\overline{N})_{b=0}}. 
\end{equation}
Linear fits to the \trento{} simulations in Fig.~\ref{fig:binomial} return $\gamma\simeq 3.5$, and we use this value throughout this paper.
Note that it is somewhat larger than that inferred from Duke~\cite{Moreland:2018gsh} and JETSCAPE parametrizations~\cite{JETSCAPE:2020mzn}, which give $\gamma\simeq 2.8$ and $\gamma\simeq 1.9$ respectively~\cite{Samanta:2023kfk}. 
This larger value implies larger centrality fluctuations in mid-central and peripheral collisions. 

This value of $\gamma=3.5$ only applies to one centrality classifier, namely, $N_{ch}$, which is the charged multiplicity seen around mid-rapidity.
We assume for simplicity that it is identical for the alternative centrality classifier, $E_T$, used by ATLAS.\footnote{
If we carry out the same study, replacing $N_{ch}$ with $E_T$ everywhere, we obtain a larger value, $\gamma\simeq 5.5$.
However, it does not seem justified to use the \trento{} model to model $E_T$ fluctuations. 
The reason is that $E_T$ is measured at large rapidities, while the \trento{} model treats the two nuclei symmetrically, which is justified only near mid-rapidity.}

\begin{figure*}[ht]
\begin{center}
\includegraphics[width=0.95\linewidth]{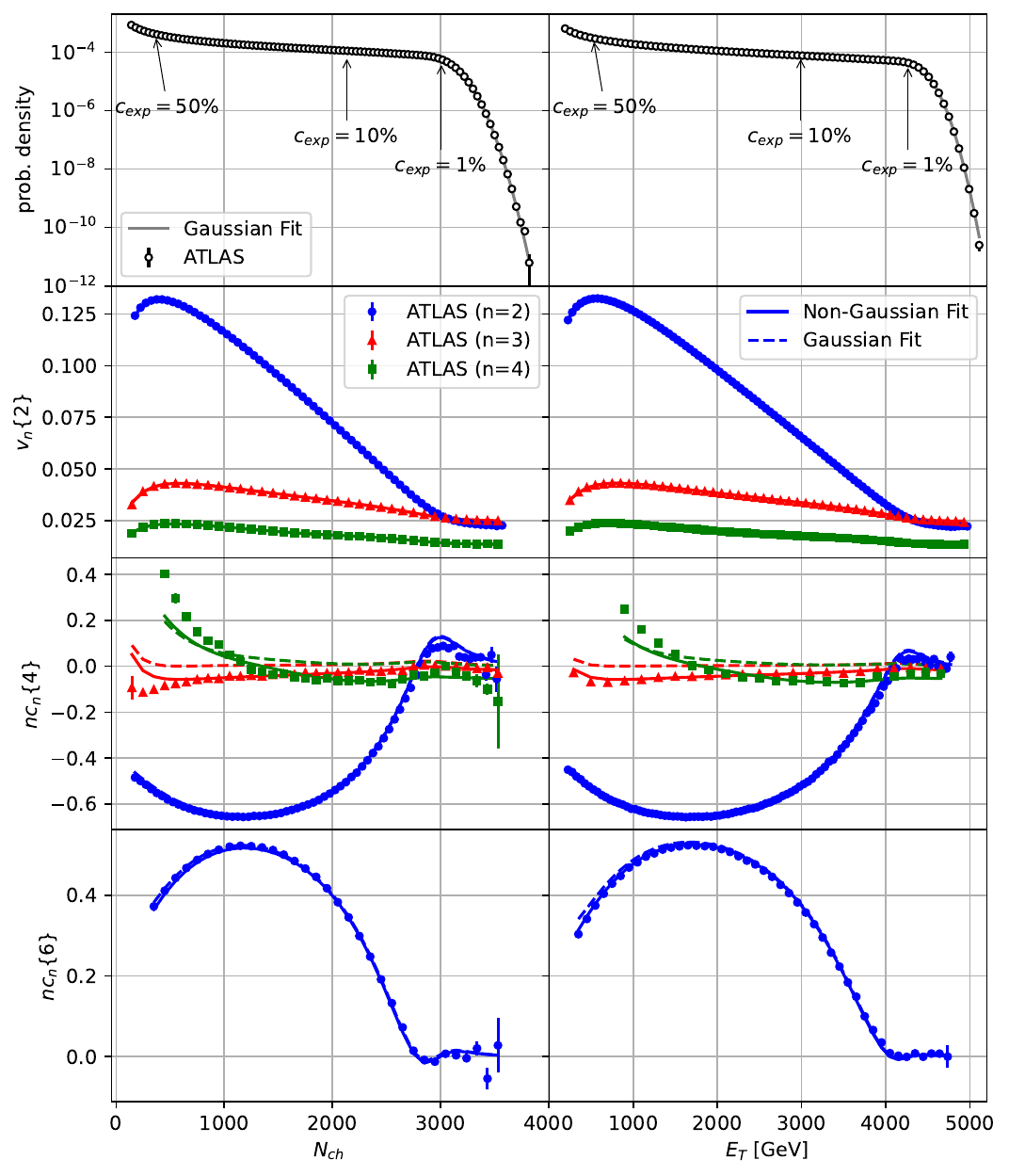} 
\end{center}
\caption{ 
  \label{fig:ATLAS}
  Top: distributions of the charged multiplicity $N_{ch}$ (left) and transverse energy $E_T$ (right) measured by the ATLAS collaboration in Pb+Pb collisions at $\sqrt{s_{NN}}=5.02$~TeV~\cite{ATLAS:2019peb}. Lines are our fits (Sec.~\ref{s:pcfixedN}). 
Lower panels display the cumulants of anisotropic flow, defined by Eqs.~(\ref{defncn}), measured in events with given $N_{ch}$ (left) or $E_T$ (right), for charged particles in the transverse momentum range $0.5<p_t<5$~GeV/$c$.
Symbols are ATLAS data (Sec.~\ref{s:data}). Solid lines and dashed lines are fits with and without non-Gaussianities (Secs.~\ref{s:procedure} and \ref{s:fitvsdata}). 
}
\end{figure*}    

\subsection{Centrality fluctuations in the ATLAS experiment}
\label{s:pcfixedN}

The measured distribution of $N$, $P(N)$,  is the integral over $c$ of the distribution at fixed $c$, $P(N)=\int_0^1 P(N|c)dc$. 
We assume that $P(N|c)$ is Gaussian, and parametrize the mean $\overline{N}(c)$ as the exponential of a polynomial of degree 5, which provides enough degrees of freedom to obtain good fits over the $0-70\%$ centrality range. 
The resulting fits to the ATLAS distributions of $N_{ch}$ and $E_T$ are displayed in the top panels of Fig.~\ref{fig:ATLAS}. 
The difference between fit and data is at the percent level for both distributions.

\begin{figure*}[ht]
\begin{center}
\includegraphics[width=.85\linewidth]{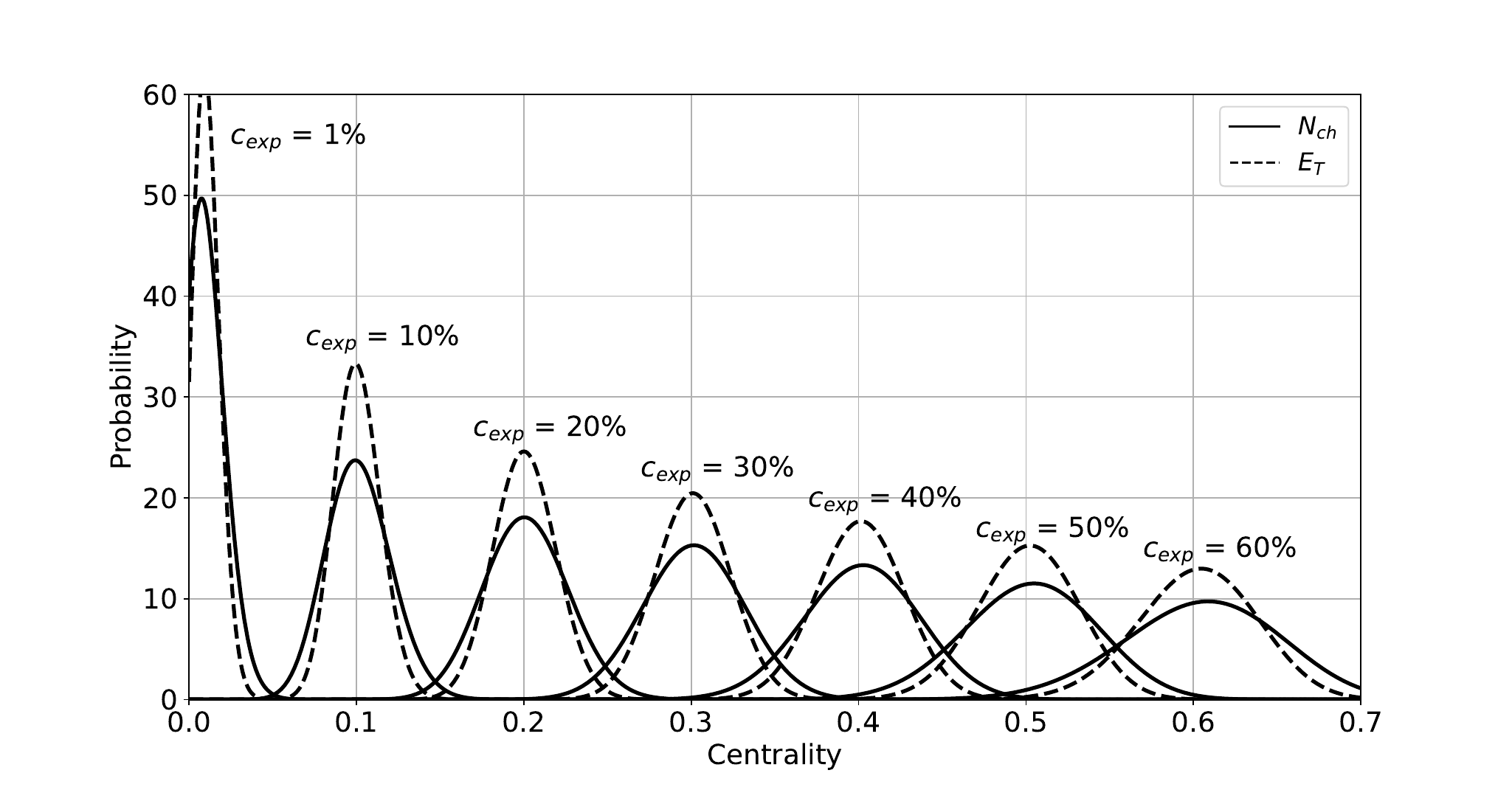}
\end{center}
\caption{ 
  \label{fig:centralityfluctuations}
Bayesian reconstruction of the probability density function of the true centrality, for several fixed values of the centrality $c_{\rm exp}$ defined experimentally using the multiplicity of charged particles (full lines) or the transverse energy in forward calorimeters (dashed lines). 
 }
\end{figure*}    

The probability distribution of $c$ at fixed $N$ is then given by Bayes' theorem $P(c|N)=P(N|c)/P(N)$. 
it is displayed in Fig.~\ref{fig:centralityfluctuations} for several values of the experimental centrality $c_{\rm exp}$.\footnote{Note that the determination of the experimental centrality requires proper calibration through an anchor point~\cite{ALICE:2013hur} which is provided by the collaborations.
Here it is assumed that $c_{\rm exp}=50\%$ corresponds to $E_T>525$~GeV.}
The width of the distribution becomes broader as the centrality percentile increases. 
It is therefore essential to assess the effect of these centrality fluctuations on $v_n$ fluctuations, which we now discuss.

\section{Fits to ATLAS data}
\label{s:fits}

\subsection{Data on anisotropic flow cumulants}
\label{s:data}

We use data from the ATLAS collaboration~\cite{ATLAS:2019peb} for charged hadrons in the pseudorapidity window $|\eta|<2.5$ and transverse momentum range $0.5<p_t<5$~GeV$/c$. 
Their  rms anisotropic flow $v_n\{2\}$ is measured for $n=2,3,4$, as well as cumulants of order 4 and 6, which are presented in the form of scaled cumulants, $nc_n\{4\}$ and $nc_n\{6\}$. 
These observables are defined by 
\begin{align}
v_n\{2\}&\equiv\langle |v_n|^2|N\rangle^{1/2}\nonumber\\
nc_n\{4\}&\equiv\frac{\langle |v_n|^4|N\rangle}{\langle |v_n|^2|N\rangle^2}-2\nonumber\\
nc_n\{6\}&\equiv\frac{\langle |v_n|^6|N\rangle}{4\langle |v_n|^2|N\rangle^3}-\frac{9\langle |v_n|^4|N\rangle}{4\langle |v_n|^2|N\rangle^2}+3,
\label{defncn}
\end{align}
where $\langle X|N\rangle$ denotes the average value of $X$ over events in a centrality class, i.e., for a fixed value of the centrality classifier $N$. 
The analysis is carried out with two different centrality classifications, $N=N_{ch}$ and  $N=E_T$. 
Results are displayed in the lower panels of Fig.~\ref{fig:ATLAS}. 
For the cumulant of order $6$, ATLAS provides results for $n=2$ and $3$, but there are only few data points with large error bars for $n=3$, and we will not use them. 

The first observation in Fig.~\ref{fig:ATLAS} is that cumulants of $v_2$ are larger and vary more steeply than that those of higher harmonics.
Before describing the details of our model calculation, we briefly recall the origin of this well-known phenomenon within the Gaussian model of fluctuations. 
Under the assumption of linear response, the distribution of $v_2$ is similar to the distribution of $\varepsilon_2$, Eq.~(\ref{gaussian}), where one replaces $\bar\varepsilon_2$ with $\bar v_2$ (the mean elliptic flow in the reaction plane), and $\sigma_{\varepsilon_2}$ with $\sigma_{v_2}$ (width of elliptic flow fluctuations). 
That of $v_3$ has the same form with $\bar v_3=0$ and  $\sigma_{v_2}$ replaced with $\sigma_{v_3}$. 
Then, neglecting centrality fluctuations, the rms $v_2$ and $v_3$ are given by: 
\begin{align}
v_2\{2\}&=(\bar v_2^2+\sigma_{v_2}^2)^{1/2}\nonumber\\
v_3\{2\}&=\sigma_{v_3}.
\label{rmsflows}
\end{align}
While the centrality dependence of the fluctuation $\sigma_{v_n}$ is mild, $\bar v_2$ decreases steeply as the collision becomes more central, which drives the variation of $v_2\{2\}$. 
Similarly, neglecting centrality fluctuations and non-Gaussianities, higher-order cumulants of $v_2$ are given by 
\begin{align}
nc_2\{4\}&=-\left(\frac{\bar v_2^2}{\bar v_2^2+\sigma_{v_2}^2}\right)^2\nonumber\\
nc_2\{6\}&=\phantom{-}\left(\frac{\bar v_2^2}{\bar v_2^2+\sigma_{v_2}^2}\right)^3.
\label{nc246}
\end{align}
They are determined by the relative magnitudes of $\bar v_2$ and $\sigma_{v_2}$. 
The changes of sign in central collisions are explained by centrality fluctuations~\cite{Zhou:2018fxx,Alqahtani:2024ejg}. 

\subsection{Fitting procedure}
\label{s:procedure}

We assume that anisotropic flow is driven by the true centrality $c$. 
Averages at fixed $N$ are computed in two steps~\cite{Alqahtani:2024ejg}. 
First, one evaluates the average at fixed centrality, then one averages over centrality fluctuations: 
\begin{equation}
\label{decomposition}
\langle |v_n|^k|N\rangle= \int_0^1 \langle |v_n|^k|c\rangle P(c|N) dc, 
\end{equation}
where $P(c|N)$ is the probability displayed in Fig.~\ref{fig:centralityfluctuations} for selected values of $N$.
This equation amounts to neglecting the correlation between $v_n$ and $N$ at fixed centrality, which has a minor effect on the reconstruction~\cite{Alqahtani:2024ejg}. 

The probability distribution of $v_n$ at fixed $c$ is modeled as detailed in Sec.~\ref{s:cumulants}. 
We now list the various quantities entering this modeling, and the corresponding fit parameters. 
The variance of the fluctuations $\sigma_{v_n}^2$ and the mean elliptic flow in the reaction plane  $\bar v_2$ are smooth functions of $c$, which we choose to parametrize by rational functions. 
The coefficients of the rational functions are our fit parameters. 
Azimuthal symmetry further requires that $\bar v_2$ vanishes for $c=0$, and this constraint is implemented in our fit~\cite{Alqahtani:2024ejg}. 
We increase the degree of the polynomials in the numerator and denominator of the rational function until the quality of the fit saturates. 

Then, there are fit parameters associated with non-Gaussian corrections, which we model in a minimal way. 
We characterize these corrections by a single parameter for each Fourier harmonic. 
We borrow the intensive cumulants of initial anisotropies from the elliptic-power distribution at large $\alpha$ and small $\varepsilon_0$ (Eqs.~(\ref{epintensive}) and Fig.~\ref{fig:epintensive}), and we use Eq.~(\ref{intensivev}) for the hydrodynamic response. 
Our default fit is thus done with the following parametrization: 
\begin{equation}
\label{intensivev2}
4\Gamma_{02}^{(v_n)}=\Gamma_{12}^{(v_n)}=\Gamma_{22}^{(v_n)}=-\frac{2}{\kappa_n^2}. 
\end{equation}
Therefore, the only fit parameter is the hydrodynamic response coefficient $\kappa_n$. 
The fluctuation asymmetry $\Gamma_{02}^{(v_n)}$  and skewness $\Gamma_{12}^{(v_n)}$ play a role only for elliptic flow, $n=2$. 

The kurtosis of elliptic flow deserves a more detailed discussion, as it varies strongly with the average anisotropy (Fig.~\ref{fig:epintensive}). 
However, its contribution is smaller than that of the skewness, except for central collisions where it is the dominant contribution (see the discussion in Sec.~\ref{s:fitvsdata} below). 
This justifies the choice (\ref{intensivev2}). 
In order to check the robustness of the result, we also carry out the fit with zero kurtosis, $\Gamma_{22}^{(v_2)}=0$. 

The moments at fixed $c$ entering the right-hand side of Eq.~(\ref{decomposition}) are evaluated as a function of $\sigma_{v_n}$, $\bar v_2$ and the intensive non-Gaussianities $\Gamma_{m2}^{(v_n)}$ using the formulas given in Appendix~\ref{s:moments}. 

Fits are carried out independently for $v_2$ and $v_3$. 
For each harmonic, we fit simultaneously all cumulant data with both centrality classifiers.
For $v_4$, the nonlinear contribution in Eq.~(\ref{nonlinear}) is evaluated by importing $v_2$ from the fit to $v_2$ data, and treating the response coefficient $\chi_{4,22}$ as an additional fit parameter, which we assume to be independent of centrality.
The linear contribution is parametrized in a way similar to $v_3$. 

\subsection{Comparison between fit and data}
\label{s:fitvsdata}

\begin{figure*}[ht]
\begin{center}
\includegraphics[width=.8\linewidth]{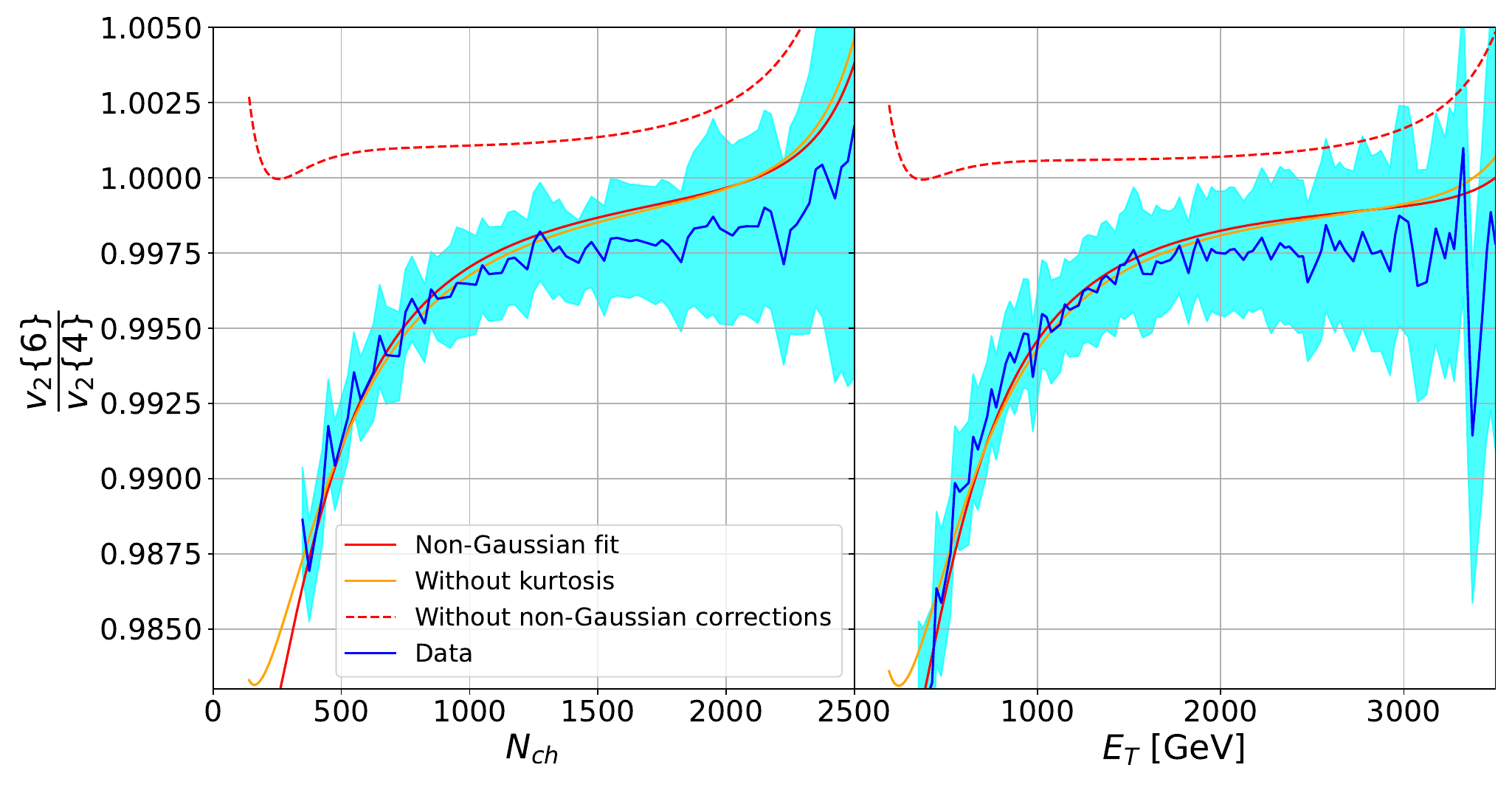} 
\end{center}
\caption{ 
  \label{fig:splitting}
 Ratio $v_2\{6\}/v_2\{4\}$ as a function of $N_{ch}$ (left) or $E_T$ (right). 
 The shaded band is the experimental error. 
 Full lines represent the fits with non-Gaussian corrections included, according to Eq.~(\ref{intensivev2}) and with the kurtosis $\Gamma_{22}^{(v_2)}$ set to 0. 
 Dashed lines are obtained by setting the non-Gaussian corrections to 0, so that the deviation from unity solely comes from centrality fluctuations. 
}
\end{figure*}    

The lower panels of Fig.~\ref{fig:ATLAS} display ATLAS data together with our fits with the Gaussian model (dashed lines) and with non-Gaussianities added (solid lines).  
The differences between the two sets of lines illustrate the effects of non-Gaussian fluctuations. 
We discuss these results from top to bottom. 
Fits to $v_n\{2\}$ data are excellent, as a function of both centrality classifiers $N_{ch}$ or $E_T$.

Next come the cumulants of order 4.
$nc_2\{4\}$ data are well reproduced. The fit goes too positive in ultracentral collisions, and we do not understand the origin of this slight discrepancy. 
$nc_3\{4\}$ is well reproduced only after adding the kurtosis. 
This can be understood in the following way. 
Neglecting centrality fluctuations, we obtain using Eqs.~(\ref{defncn}), (\ref{momentsv3}) and (\ref{intensivev2})
\begin{equation}
  \label{v3kurtosis}
nc_3\{4\}=\sigma_{v_3}^2\Gamma_{22}^{(v_3)}=-\frac{2v_3\{2\}^2}{\kappa_3^2}. 
\end{equation}
This equation shows that  the negative $nc_3\{4\}$ is a direct signature of the negative kurtosis $\Gamma_{22}^{(v_3)}$,  as already mentioned in Sec.~\ref{s:ellipticpower}. 
Note that a centrality-independent $\Gamma_{22}^{(v_3)}$ captures most of the centrality dependence of $nc_3\{4\}$, which illustrates the mild system-size dependence of intensive cumulants. 
The fit underpredicts the data, in absolute value, for the most peripheral collisions. 
The natural explanation here is that we have assumed a constant hydrodynamic response coefficient $\kappa_3$. 
However, due to viscous damping,  $\kappa_3$ is smaller in peripheral collisions~\cite{Gardim:2020mmy}, as will be discussed in Sec.~\ref{s:results}, and this implies a larger $|nc_3\{4\}|$. 
This still does not explain why $nc_3\{4\}$ goes more negative as a function of $N_{ch}$ than as a function of $E_T$.
We do not have any explanation of this difference. 

For $v_4$, the cumulant of order 4 is also driven by the kurtosis $\Gamma_{22}^{(v_4)}$ in central collisions. 
In mid-central collisions, on the other hand, the cumulant is largely driven by the non-linear response, and by the fluctuations of $v_2$~\cite{Giacalone:2016mdr}. 
While our fit captures the change of sign of $nc_4\{4\}$, it underpredicts its value for peripheral collisions, both as a function of $N_{ch}$ and $E_T$.
This is a hint that modeling $v_4$ as the sum of two independent components, as in Eq.~(\ref{nonlinear}), may be an oversimplification. 

Finally, the cumulant of order 6, which we only display for $v_2$, is described with good accuracy only after non-Gaussian corrections are added. 
The effect of these corrections is more clearly seen by plotting the ratio $v_2\{6\}/v_2\{4\}$, which is displayed in Fig.~\ref{fig:splitting}.
It is very close to unity, because $v_2\{4\}$ and $v_2\{6\}$ both coincide with $\bar v_2$, up to centrality fluctuations and non-Gaussian corrections. 
The dotted line displays the effect of centrality fluctuations. 
Despite the significant centrality spread observed in Fig.~\ref{fig:centralityfluctuations}, they have a modest effect, and increase the ratio $v_2\{6\}/v_2\{4\}$ relative to unity. 
This implies that a value smaller than unity can solely be attributed to non-Gaussianities.
In order to understand their effect, we derive an approximate expression of $v_2\{6\}/v_2\{4\}$, by neglecting centrality fluctuations, using Eqs.~(\ref{defncn}) and (\ref{momentsv2}), and linearizing in $\Gamma_{m2}^{(v_2)}$:
\begin{align}
\frac{v_2\{6\}}{v_2\{4\}}&\equiv 
\frac{nc_n\{6\}^{1/6}}{(-nc_n\{4\})^{1/4}}\nonumber\\
&=1+\frac{\sigma_{v_2}^4}{4\bar v_2^2}\Gamma_{12}^{(v_2)}
+\frac{3\sigma_{v_2}^6}{4\bar v_2^4}\Gamma_{22}^{(v_2)}. 
\label{v4v6splitting}
\end{align}
First, one notes that the splitting between $v_2\{4\}$ and $v_2\{6\}$ only involves the skewness and the kurtosis, not the fluctuation asymmetry $\Gamma_{02}^{(v_2)}$.\footnote{To linear order in non-Gaussianities, the splitting between $v_2\{6\}$ and higher-order cumulants such as $v_2\{8\}$ and $v_2\{10\}$~\cite{CMS:2023bvg}  only depends on the skewness $\Gamma_{12}^{(v_2)}$, not on the kurtosis $\Gamma_{22}^{(v_2)}$.}
This implies that $\Gamma_{02}^{(v_2)}$ cannot be constrained from data.\footnote{Note, however, that a negative  $\Gamma_{02}^{(v_2)}$, which is expected on general grounds, lowers the mean elliptic flow in the reaction plane $\bar v_2$ relative to $v_2\{4\}$ and $v_2\{6\}$.}
The contribution of the skewness dominates over that of the kurtosis because $\bar v_2>\sigma_{v_2}$, except for central collisions.
Therefore, the observation that $v_2\{6\}<v_2\{4\}$ is a direct consequence of the negative skewness~\cite{Giacalone:2016eyu,CMS:2017glf}. 

Fig.~\ref{fig:splitting} shows that the ratio $v_2\{6\}/v_2\{4\}$ is very well reproduced by our fit.
We also display the fit obtained by setting the kurtosis to zero, which is close to the default fit.
A constant intensive skewness suffices to capture the whole centrality dependence, in contrast to the usual standardized skewness which depends significantly on centrality~\cite{CMS:2017glf,ALICE:2018rtz}. 
This again illustrates the usefulness of intensive cumulants. 

\section{Results and interpretation}
\label{s:interpretation}

\subsection{Reconstructed centrality dependence}
\label{s:results}

\begin{figure}[ht]
\begin{center}
\includegraphics[width=.8\linewidth]{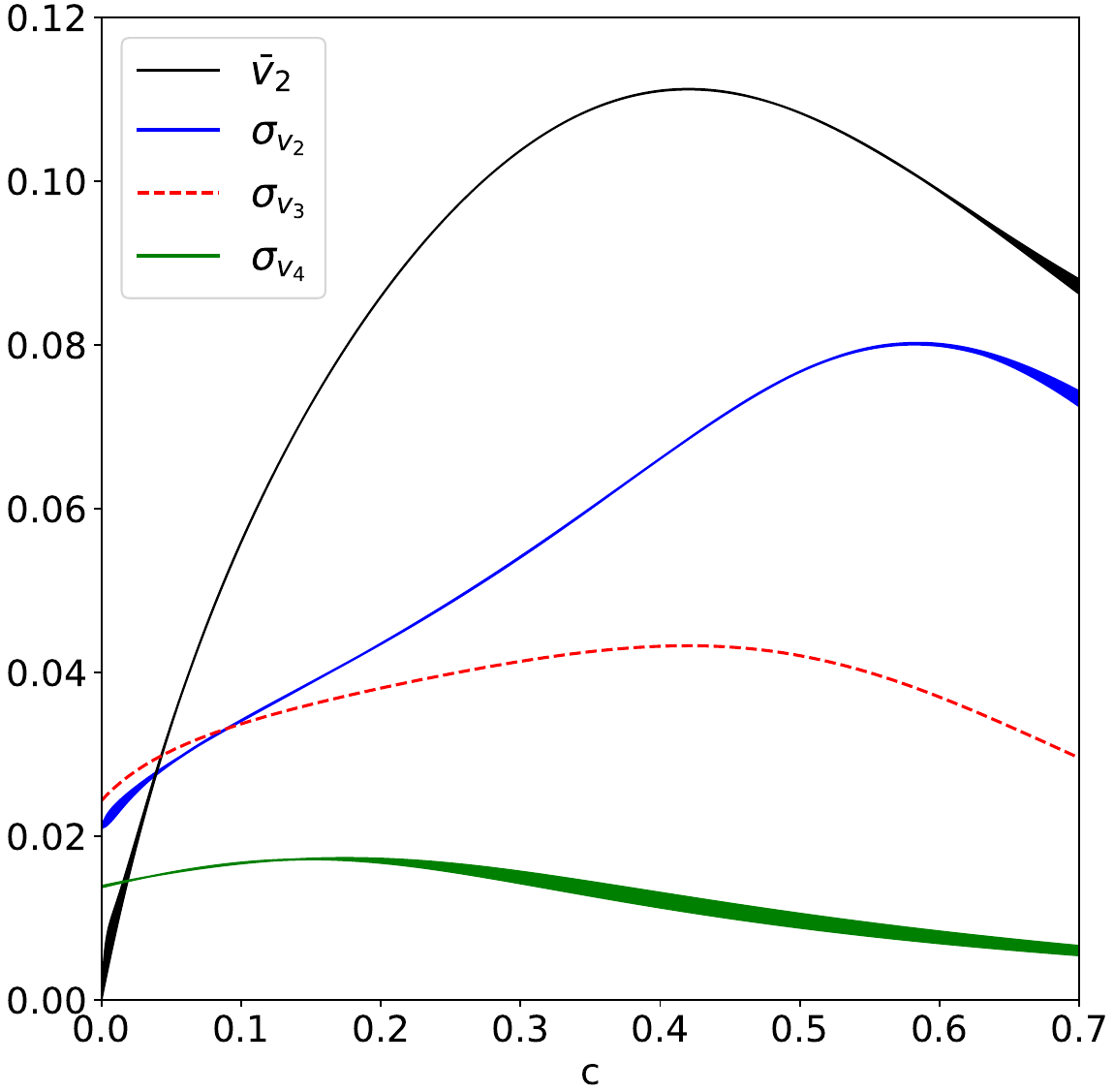} 
\end{center}
\caption{ 
  \label{fig:reconstructed}
  Bayesian reconstruction of the centrality dependence of the mean elliptic flow in the reaction plane $\bar v_2$ and of the width of flow fluctuations $\sigma_{v_n}$ for $n=2,3,4$.
  Bands for $n=2$ and $n=4$ indicate our estimate of the error (see text). 
}
\end{figure}    
Fig.~\ref{fig:reconstructed} displays the reconstructed centrality dependence of  $\bar v_2$ and $\sigma_{v_n}$.
These functions are our fit parameters (recall that $\bar v_2$  and $\sigma_{v_n}^2$ are parametrized as rational functions of $c$). 
Triangular flow is the simplest, since $\sigma_{v_3}=v_3\{2\}$ after unfolding centrality fluctuations. 
For $v_2$, one must separate the contributions of $\bar v_2$ and $\sigma_{v_2}$. 
Neglecting centrality fluctuations and non-Gaussianities, $\bar v_2\simeq v_2\{4\}$ and $\sigma_{v_2}\simeq(v_n\{2\}^2-v_n\{4\}^2)^{1/2}$. 
For $v_4$, $\sigma_{v_4}$ represents the linear part, sometimes denoted by $v_{4L}$, which the ALICE collaboration has isolated using a different method~\cite{ALICE:2017fcd}.\footnote{They extract the nonlinear response coefficient $\chi_{4,22}$ from the correlation between $v_4$ and $v_2$, then subtract the nonlinear contribution from $v_4\{2\}$. It is interesting to note that their result for $v_{4L}$ has a centrality dependence similar to our $\sigma_{v_4}$.}
The error bands for $v_2$ and $v_4$ are obtained by adding more fit parameters than necessary (i.e., increasing the degree of the polynomial in the numerator and denominator of the rational functions). 
This does not improve significantly the quality of the fit, but results in small changes of the reconstructed centrality dependence. 
Note that the error band becomes broader when $c$ approaches the upper limit $0.7$. 
This is natural since we only include events with $c_{\rm exp}<0.7$, and $c$ fluctuates with respect to $c_{\rm exp}$ so that not all events with $c<0.7$ are included.  
For $\sigma_{v_3}$, the fit algorithm fails to converge when we increase the number of fit parameters, so that we are unable to obtain a meaningful error band with this method, but we expect a smaller error than for $\sigma_{v_2}$ and $\sigma_{v_4}$ anyway. 

The mean elliptic flow in the reaction plane $\bar v_2$ first increases linearly with centrality, then levels off and reaches a maximum slightly above 40\% centrality.
This coincides with the experimental centrality at which $v_2\{4\}$ reaches its maximum~\cite{ALICE:2016ccg,CMS:2023bvg}.
Therefore, centrality fluctuations and non-Gaussian corrections, which cause $v_2\{4\}$ to deviate from $\bar v_2$, have a small effect.

In hydrodynamics, $\sigma_{v_n}$  is a linear response to $\sigma_{\varepsilon_n}$ (or $\sigma_{{\cal C}_4}$ for $n=4$), which 
increases monotonically as a function of impact parameter~\cite{Niemi:2015qia}, since fluctuations are larger in smaller systems.
Therefore, a decrease of $\sigma_{v_n}$ can safely be attributed to a decrease of the response coefficient $\kappa_n$ as a function of centrality.
Such a decrease is expected as a consequence of viscous damping, which decreases $\kappa_n$ by a larger fraction as $c$ increases, as a consequence of Reynolds number scaling~\cite{Gardim:2020mmy}.
The larger the viscosity, the earlier the decrease occurs as a function of centrality. 
Since viscous damping increases with $n$ like $n^2$~\cite{Teaney:2012ke}, one expects that $\sigma_{v_n}$ decreases earlier for larger $n$. 
This is confirmed by our results in Fig.~\ref{fig:reconstructed}, where $\sigma_{v_n}$ reaches its maximum around $60\%$ centrality for $n=2$, $40\%$ centrality for $n=3$, and below $20\%$ centrality for $n=4$.

\subsection{Hydrodynamic response coefficients} 
\label{s:coefficients}

\begin{figure}[ht]
\begin{center}
\includegraphics[width=.7\linewidth]{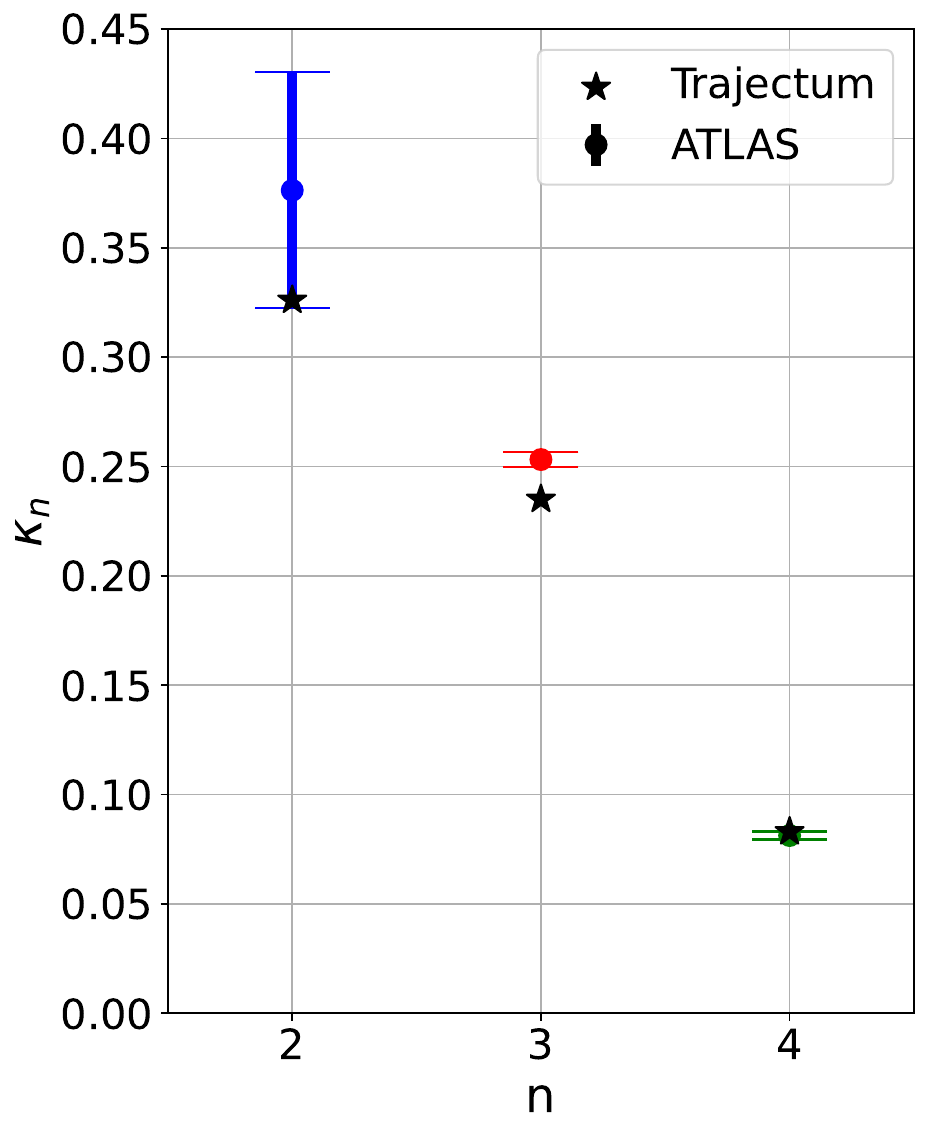} 
\end{center}
\caption{ 
  \label{fig:coefficients}
Hydrodynamic response coefficients inferred from intensive non-Gaussianities of $v_n$ fluctuations.  
The error bar for $\kappa_2$ is obtained by repeating the fit with the kurtosis $\Gamma_{22}^{(v_2)}$ set to zero (see text).
Symbols are direct estimates of response coefficients from hydrodynamic simulations~\cite{Nijs:2020roc}.}
\end{figure}    

The other fit parameters are the intensive non-Gaussianities in each Fourier harmonic.
As explained in Sec.~\ref{s:procedure}, we assume that the intensive non-Gaussianities of initial anisotropies are those of the elliptic-power distribution, so that the measured intensive non-Gaussianities of anisotropic flow yield direct estimates of hydrodynamic linear response coefficients through Eq.~(\ref{intensivev2}).\footnote{Note that these are effective values involving an averaging over centrality, since we neglect the centrality dependence of $\kappa_n$ observed in Sec.~\ref{s:results}.}
Fig.~\ref{fig:coefficients} displays their values as a function of $n$. 
$\kappa_3$ and $\kappa_4$ are compatible with the values inferred from central collisions alone~\cite{Alqahtani:2024ejg}. 
The estimate of $\kappa_2$ is  new. It is driven by the small splitting between $v_2\{6\}$ and $v_2\{4\}$, which is determined accurately in mid-central collisions.
Its value depends on whether the fit is done with the kurtosis  $\Gamma_{22}^{(v_2)}$ taken from Eq.~(\ref{intensivev2}), or with $\Gamma_{22}^{(v_2)}=0$   (see Fig.~\ref{fig:splitting}), hence the large error bar. 
The lower value corresponds to $\Gamma_{22}^{(v_2)}=0$. 

We now compare these data-driven estimates of $\kappa_n$ with hydrodynamic simulations, where initial anisotropies $\varepsilon_n$ and the anisotropic flow coefficients $v_n$ can be calculated on an event-by-event basis. 

For $n=2$ and $n=3$, 
$\kappa_n$ can then obtained directly from Eq.~(\ref{linear}), either on an event-by-event basis~\cite{Ambrus:2024hks,Ambrus:2024eqa}, or as the ratio of the mean $|v_n|$ to the mean $|\varepsilon_n|$~\cite{Qiu:2011hf}, or as the slope of the linear best fit~\cite{Nijs:2020roc}.
Qiu {\it et al.\/}~\cite{Qiu:2011hf} report $\kappa_2\simeq 0.22$ and  $\kappa_3\simeq 0.18$ with a viscosity over entropy ratio $\eta/s=0.2$ for particles in the transverse momentum range $0.2<p_t<5$~GeV/c, which is the selection used by the ALICE experiment~\cite{ALICE:2011ab}. 
For the $p_t$ selection of the ATLAS experiment, $0.5<p_t<5$~GeV/c, $v_2$ and $v_3$ are larger by a factor $\simeq 1.4$,  and the corresponding values are $\kappa_2\simeq 0.31$ and  $\kappa_3\simeq 0.26$. 
Both are in fair agreement with our estimates. 
The more recent Trajectum analysis~\cite{Nijs:2020roc}, whose parameters are fitted to LHC data, gives for the 20\%-40\% centrality window $\kappa_2=0.326$ and $\kappa_3=0.235$, with the same cuts as for the ATLAS expriment, again in agreement with our results. 

Determining $\kappa_4$ in event-by-event hydrodynamics is more difficult, because the separation between the linear and the nonlinear response terms in Eq.~(\ref{nonlinear}) is well defined only if one specifies the value of $\chi_{4,22}$. 
One can resort to hydrodynamic simulations of ultra-central collisions, where the nonlinear response is negligible and ${\cal C}_4\simeq\varepsilon_4$, so that $\kappa_4$ can again be defined as $v_4/\varepsilon_4$. 
One finds in these simulations~\cite{Luzum:2012wu} that $\varepsilon_4$ is typically of the same magnitude as $\varepsilon_3$, while $v_4$ is smaller by a factor $\sim 2$, corresponding to a $\kappa_4$ smaller than $\kappa_3$ by a factor $\sim 2$, larger than our estimate. 
On the other hand, the Trajectum calculation~\cite{Nijs:2020roc} gives the value $\kappa_4=0.083$ for the slope of $v_4$ versus $\varepsilon_4$, albeit for mid-central collisions, where one does not expect a simple linear dependence. 
This value is in excellent agreement with ours (Fig.~\ref{fig:coefficients}). 

It is striking that our purely data-driven approach, where the magnitude of initial anisotropies is not specified, returns reasonable values of linear response coefficients, solely by scrutinizing the non-Gaussianity of flow fluctuations. 

Finally, the value of the nonlinear response coefficient returned by our fit to ATLAS data is  $\chi_{4,22}=0.98\pm 0.01$, very close to unity. 
The value obtained by ALICE and CMS by analyzing the correlation between $v_2$ and $v_4$ is $\chi_{4,22}\simeq 1.1$ in mid-central collisions~\cite{ALICE:2016kpq,CMS:2019nct,ALICE:2020sup}, for which the relative contribution of the nonlinear response is largest. 
This is close to our estimate, but a precise comparison should take into account the different $p_t$ cuts. 

\section{Conclusions}

We have obtained a data-driven estimate of centrality (or impact parameter) fluctuations in Pb+Pb collisions up to 70\% centrality. 
The magnitude of these fluctuations was already known for central collisions. 
We have achieved a robust extrapolation to all centralities, which is inspired by models of initial conditions. 
These models show that fluctuations are significantly reduced in central collisions, because the number of participant nucleons is close to its maximum number. 
This in turn implies that centrality fluctuations are larger in semi-central collisions than one may have thought: 
For an experimentally-determined centrality of 50\%, the distribution of the true centrality spans a broad range from 40\% to 60\%, with a full width at half-maximum of $\sim 10\%$. 

We have unfolded the effect of these centrality fluctuations on anisotropic flow, and reconstructed the probability distribution of $v_n$ at fixed impact parameter. 
For $v_2$ and $v_3$, we have parametrized this distribution as a Gaussian, up to small non-Gaussian corrections originating from the condition that initial anisotropies $\varepsilon_n$, which generate anisotropic flow, are smaller than unity. 
For $v_4$, we have assumed a simple superposition of a nonlinear response induced by $v_2$, and an additional component due to fluctuations. 
We have introduced intensive measured of non-Gaussianities (skewness and kurtosis) which depend little on centrality. 

Our fit to ATLAS data is good in general, but with degraded quality in the most peripheral collisions, in particular for $v_4$.
The estimates of intensive non-Gaussianities translate into data-driven estimates of hydrodynamic response coefficients, which are in good agreement with hydrodynamic calculations. 
These estimates are based on two assumptions: 
That the intensive non-Gaussianities of initial anisotropy fluctuations are universal, and that anisotropic flow is driven by simple response relations to initial anisotropies. 
These assumptions, which are summarized by Eq.~(\ref{intensivev2}), can be checked by carrying out event-by-event hydrodynamic simulations at fixed impact parameter. 

It will also be useful to extend this study to mixed cumulants involving several harmonics, which have also been analyzed by ATLAS. 
In particular, including data on the correlation between $v_2$ and $v_4$ should improve our understanding of $v_4$. 

\begin{acknowledgments}
We thank Giuliano Giacalone for useful comments on the manuscript. 
A.K. is supported by
the U.S. Department of Energy, Office of Science, Office
of Nuclear Physics, grant No. DE-FG02-05ER41367.
\end{acknowledgments}

\appendix

\section{Relation with previous cumulant expansions}
\label{s:cversusk}

We have defined the cumulants of eccentricity $c_{mp}$ in Eq.~(\ref{defcmp}) in terms of the complex eccentricity $\varepsilon_2$ and its complex conjugate $\varepsilon_2^*$, as proposed by Mehrabpour~\cite{Mehrabpour:2020wlu}. 
Previous works~\cite{Abbasi:2017ajp,Bhalerao:2018anl,CMS:2023bvg} have used instead the cumulants $\kappa_{mp}$ of the real and imaginary parts  $\varepsilon_x$ and $\varepsilon_y$:
\begin{equation}
\kappa_{mp}\equiv \langle (\varepsilon_x)^m(\varepsilon_y)^p\rangle_c,  
\end{equation}
where symmetry with respect to the reaction plane implies that $\kappa_{mp}$ vanishes for odd $p$.  
The relations between the two sets of cumulants~\cite{Alqahtani:2024ejg} are obtained by equating their generating functions~\cite{Mehrabpour:2020wlu}:
\begin{equation}
\label{gencumulants}
\sum_{m,p\ge 0}\frac{(\lambda_x+i\lambda_y)^m(\lambda_x-i\lambda_y)^{p}}{m!p!}c_{mp}=
\sum_{m,p\ge 0}\frac{2^{m+p}\lambda_x^m\lambda_y^{p}}{m!p!}\kappa_{mp}.
\end{equation}
The explicit expressions of $\kappa_{mp}$ as a function of $c_{mp}$ are: 
\begin{align}
\kappa_{10}&= c_{01}\nonumber\\
\kappa_{20}&= \frac{1}{2}c_{11}+\frac{1}{2}c_{02}\nonumber\\
\kappa_{02}&= \frac{1}{2}c_{11}-\frac{1}{2}c_{02}\nonumber\\
\kappa_{30}&= \frac{3}{4}c_{12}+\frac{1}{4}c_{03}\nonumber\\
\kappa_{12}&= \frac{1}{4}c_{12}-\frac{1}{4}c_{03}\nonumber\\
\kappa_{40}&= \frac{3}{8}c_{22}+\frac{1}{2}c_{13}+\frac{1}{8}c_{04}\nonumber\\
\kappa_{22}&= \frac{1}{8}c_{22}-\frac{1}{8}c_{04}\nonumber\\
\kappa_{04}&= \frac{3}{8}c_{22}-\frac{1}{2}c_{13}+\frac{1}{8}c_{04}\nonumber\\
\kappa_{50}&= \frac{5}{8}c_{23}+\frac{5}{16}c_{14}+\frac{1}{16}c_{05}\nonumber\\
\kappa_{32}&= \frac{1}{8}c_{23}-\frac{1}{16}c_{14}-\frac{1}{16}c_{05}\nonumber\\
\kappa_{14}&= \frac{1}{8}c_{23}-\frac{3}{16}c_{14}+\frac{1}{16}c_{05},
\end{align}
and so on. 
The terms in the right-hand side are ranked by decreasing magnitude, the second term in each row being smaller by a factor $\sim c_{01}^2=\bar\varepsilon_2^2$ than the first term, and so on.  
The approximate relations $\kappa_{12}\approx \frac{1}{3}\kappa_{30}$, $\kappa_{22}\approx \frac{1}{3}\kappa_{40}$, $\kappa_{32}\approx\kappa_{14}\approx  \frac{1}{5}\kappa_{50}$,  which have been derived using the elliptic-power distribution~\cite{CMS:2023bvg} are in fact more general. 
They are obtained by keeping only the leading term, which is justified for a mild breaking of azimuthal symmetry. 

\section{Moments of the distribution of $|v_n|$ at fixed centrality}
\label{s:moments}

The moments can be expressed as a function of the cumulants using the formalism of generating functions, as detailed in Appendix~A of Ref.~\cite{Alqahtani:2024ejg}. 
This calculation is straightforward using formal calculus. 
We list the results which are relevant for this paper. 
For $v_2$, we need the moments up to order $6$: 
\begin{widetext}
\begin{align}
  \langle |v_2|^2|c\rangle&= \sigma_{v_2}^2+\bar v_2^2\nonumber\\
  \langle |v_2|^4|c\rangle&=\left(2+\sigma_{v_2}^2\Gamma_{22}^{(v_2)} \right)\sigma_{v_2}^4+4\left(1+\sigma_{v_2}^2\Gamma_{12}^{(v_2)}\right) \sigma_{v_2}^2\bar v_2^2+
  \bar v_2^4\left( 1+\sigma_{v_2}^2\Gamma_{02}^{(v_2)} \right)^2  \nonumber\\
  \langle |v_2|^6|c\rangle&=\left(6+9 \sigma_{v_2}^2\Gamma_{22}^{(v_2)}\right)\sigma_{v_2}^6+9 \left(2+\sigma_{v_2}^2(\Gamma_{22}^{(v_2)}+\Gamma_{12}^{(v_2)}(4+\sigma_{v_2}^2\Gamma_{12}^{(v_2)})\right)
  \sigma_{v_2}^4\bar v_2^2
  \nonumber\\ &+9
  \left(1+\sigma_{v_2}^2\Gamma_{02}^{(v_2)} \right)
  \left(1+\sigma_{v_2}^2(\Gamma_{02}^{(v_2)}+2\Gamma_{12}^{(v_2)}) \right)
  \sigma_{v_2}^2\bar v_2^4   +\bar v_2^6\left( 1+3\sigma_{v_2}^2\Gamma_{02}^{(v_2)} \right)^2.
  \label{momentsv2}
\end{align}
For $v_3$, we only need the moments up to order 4. 
Replacing $v_2$ with $v_3$ in Eq.~(\ref{momentsv2}) and setting $\bar v_2=0$, one obtains: 
\begin{align}
  \langle |v_3|^2|c\rangle&= \sigma_{v_3}^2\nonumber\\
  \langle |v_3|^4|c\rangle&=\left(2+\sigma_{v_3}^2\Gamma_{22}^{(v_3)}\right)\sigma_{v_3}^4.
  \label{momentsv3}
\end{align}
The moments of $v_4$ are obtained from Eq.~(\ref{nonlinear}), where the two terms are assumed to be independent: 
\begin{align}
  \langle |v_4|^2|c\rangle&=\sigma_{v_4}^2+\chi^2\langle |v_2|^4|c\rangle \nonumber\\
  \langle |v_4|^4|c\rangle&=\left(2+\sigma_{v_4}^2\Gamma_{22}^{(v_4)}\right)\sigma_{v_4}^4+4
  \chi^2\langle |v_2|^4|c\rangle\sigma_{v_4}^2+\chi^4\langle |v_2|^8|c\rangle. 
\end{align}
The value of $\langle |v_2|^4|c\rangle$ is taken from Eq.~(\ref{momentsv2}). 
The moment of order 8, $\langle |v_2|^8|c\rangle$, is obtained in a similar way, but its expression is lengthy and we do not write it. 
\end{widetext}

\end{document}